# The Google Scholar Experiment: how to index false papers and manipulate bibliometric indicators

---

**Emilio Delgado López-Cózar, Nicolás Robinson-García***

EC3 Research Group: Evaluación de la Ciencia y la Comunicación Científica, Facultad de Comunicación y Documentación, Colegio Máximo de Cartuja s/n, Universidad de Granada, 18071, Granada, Spain.

Telephone: +34 958 240920

Email addresses: edelgado@ugr.es; elrobin@ugr.es

**Daniel Torres-Salinas**

EC3 ResearchGroup: Evaluación de la Ciencia y la Comunicación Científica, Centro de Investigación Médica Aplicada, Universidad de Navarra, 31008, Pamplona, Spain.

Email address: torressalinas@gmail.com

**Abstract**

Google Scholar has been well received by the research community. Its promises of free, universal and easy access to scientific literature as well as the perception that it covers better than other traditional multidisciplinary databases the areas of the Social Sciences and the Humanities have contributed to the quick expansion of Google Scholar Citations and Google Scholar Metrics: two new bibliometric products that offer citation data at the individual level and at journal level. In this paper we show the results of a experiment undertaken to analyze Google Scholar's capacity to detect citation counting manipulation. For this, six documents were uploaded to an institutional web domain authored by a false researcher and referencing all the publications of the members of the EC3 research group at the University of Granada. The detection of Google Scholar of these papers outburst the citations included in the Google Scholar Citations profiles of the authors. We discuss the effects of such outburst and how it could affect the future development of such products not only at individual level but also at journal level, especially if Google Scholar persists with its lack of transparency.

**Keywords:** Google Citations; Google Scholar Metrics; Scientific journals; Scientific fraud; Citation counting; Bibliometrics; H-index; Evaluation; Researchers; Citation manipulation

*To whom all correspondence should be addressed.

## Introduction

When Google Scholar (hereafter GS) irrupted within the academic world it was very well-received by researchers. It offered free, universal access, using a simple and easy-to-use interface, to all scholarly documents available under an academic domain in the web and covered document types, languages and fields which were misrepresented in the main multidisciplinary scientific database at the time; Thomson Reuters' Web of Science (Kousha & Thelwall, 2011; Harzing & van der Wal, 2008). In less than ten years of existence, GS has positioned itself as one of the main information sources for researchers (Nicholas, Clark, Rowlands & Jamali, 2009; Brown & Swan, 2007; Joint Information Systems Committee, 2012). The praised capabilities of Google's algorithm to retrieve pertinent information along with the popularity of the company seem to have been inherited by GS (Walters, 2011). However, not all of its success is due to their doing; also the deep changes overcoming in scholarly communication such as Open Access or Data Sharing movements have become strong allies on such success, benefiting all parties. For instance, on the one hand GS has given repositories the visibility they were lacking of (Markland, 2006) and on the other hand, these have provided GS with a unique content no other scientific database has; preprints, theses, etc. (Kousha & Thelwall, 2007), making it a valuable resource.

In 2011, GS took a major step signaling its intentions towards research evaluation and launched the GS Citations, which offers citation profiles for researchers (Butler, 2011), and one year after GS Metrics, which offer journal rankings according to their h-index for publications from the last five years (Delgado López-Cózar & Cabezas-Clavijo, 2012). The inclusion of these tools has popularized even more the use of bibliometrics: awakening researchers' ego (Cabezas-Clavijo & Delgado López-Cózar, 2013; Wouters & Costas, 2012) and challenging the minimum requirements demanded by bibliometricians to rely on any data source for bibliometric analysis (Aguillo, 2012). Such attraction may be explained by the need researchers have to respond to ever-more demanding pressures to prove their impact in order to obtain research funding or progress in their academic career, especially in the fields of the Social Sciences and Humanities fields who see in these products a solution to their long neglected visibility in the traditional databases (Hicks, 2004).

Soon, bibliometricians turned their interest into this database and many studies emerged analyzing the possibilities of using such database for bibliometric purposes (i.e., Meho & Yang, 2007; Aguillo, 2012; Torres-Salinas, Ruiz-Perez & Delgado López-Cózar, 2009). But mainly, these studies have criticized these tools due to the inconsistencies on citation counting (Bar-Ilan, 2008), metadata errors (Jacsó, 2011) and the lack of quality control (Aguillo, 2012). These limitations are also present in their by-products. But the main reservation when considering GS and its by-products for research evaluation has to do with the lack of transparency (Wouters & Costas, 2012). This is an important limitation as it does not allow us to certify that the information offered is in fact correct, especially when trying to detect or interpret strange bibliometric patterns. GS automatically retrieves, indexes and stores any type of scientific material uploaded by an author without any previous external control; meaning that any individual or collective can modify their output impacting directly on their bibliometric performance. Hence, GS' new products project a future landscape with ethical and sociological dilemmas that may entail serious consequences in the world of science and research evaluation.

The inclusion of bibliometric tools applied in an uncontrolled environment as GS proposes, has led to another type of critical studies experimenting on their capacity to discern academic content from faked content. At this point we must refer to the study undertaken by Labbé (2010) and his inexistent researcher Ike Antkare who proved how easily computer generated tools for research evaluation can be manipulated. In similar studies, Beel, Gipp & Wilde (2010) and Beel & Gipp (2010) tested different procedures with which to influence GS' results and obtain higher ranking positions and hence, more visibility. Among others, they made use of the *Scigen* software (http://pdos.csail.mit.edu/scigen/) to create fake papers, they included modifications of previously published papers adding new references, as well as duplicates of other papers. Such papers alerted on the ease to manipulate the GS search engine. Meaning that such threats had already been denounced before the launch of GS Citations and GS Metrics. Although malpractices have also been reported in other databases due to the inclusion of the so-called 'predatory publishers' (Harzing, 2012) or simply by manipulating journals' Impact Factor (Opatrný, 2008), the lack of any type of control or transparency of GS is certainly worrying as this tool is becoming more and more popular within the research community (Bartneck & Kokkelmans, 2011).

In this paper we report our main findings and conclusions after conducting an experiment to analyze GS and its by-products' capabilities to detect manipulation in its most rudimentary version. We focus on the effects it has on GS Citations and warn of the implications such attitudes could have on GS Metrics' rankings, as this type of behavior by which someone modifies its output and impact through intentional and unrestrained self-citation is not uncommon (see e.g., Oransky, 2012). Therefore our aim is to demonstrate how easily anyone can manipulate Google Scholar's tools. We will not emphasize the technical aspects of such gaming, but its consequences in terms of research evaluation, focusing on the enormous temptation these tools can have for researchers and journals' editors, urged to increase their impact. In order to do so, we will show how the GS Citations profiles of researchers can be modified in the easiest way possible: by uploading faked documents on our personal website citing the whole production of a research group. No software program is needed, you only need to copy and paste the same text over and over again and upload the resulting documents in a webpage under an institutional domain. We will also analyze Google's capacity to detect retracted documents and delete their bibliographic records along with the citations they make. This type of studies, challenging a given system to detect errors or flaws are common in the scientific literature. This is the case for instance, of the classical studies of Peters & Ceci (1982) or Sokal (1996) criticizing the peer review system and its incapacity to detect fraud in science.

This paper is structured as follows. Firstly we described the methodology followed; how were the false documents created and where were they uploaded. Then we briefly describe our target-service which was GS Citations. Secondly, we show the effect the inclusion of false documents had on the bibliometric profiles of the researchers who received the citations. We expose the reaction of GS after such manipulation was made public and we discuss on the consequences GS' lack of transparency and easiness to manipulate may have if used as a research evaluation tool. Finally we express some concluding remarks.

## Material and methods

The main goal of this experiment was to analyze the difficulty of including false documents and citations in GS and the consequences such actions have on its by-product GS Citations and the how it would have affected GS metrics if updated at the time.

GS Citations was launched in the summer of 2011 (Butler, 2011) and made publicly available to all users in November of that same year. It has greatly expanded since, estimating in less than a year a total population of up to 72,579 users according to Ortega & Aguillo (2013). It adopts a micro-level approach offering each author their own research profile according to the contents derived from GS (Cabezas-Clavijo & Delgado López-Cózar, 2013). Authors can create their own profile by signing in and including an institutional email address. Then, GS Citations automatically assigns documents retrieved from GS to the user who can edit their bibliographic information, merge duplicates, include omitted papers or remove wrong papers. It offers the total number of citations to each document according to GS, ranking the output by times cited or publication year. Finally it includes a histogram of citations received by year and several bibliometric indicators (Total number of citations, h-index and i10 index). Also, authors can select their own keywords to label their research field, allowing to visualize research fields not as classified by a third party as it occurs in other databases, but as seen from researchers' own perspective (Ortega & Aguillo, 2012).

In order to manipulate citation counting, we adopted the most simple and rudimentary strategy we could think of. We selected the authors of this paper (Emilio Delgado López-Cózar, Nicolás Robinson-García and Daniel Torres-Salinas, hereafter EDLC, NRG and DTS) and the rest of the members of our own research group (EC3 – Evaluación de la Ciencia y de la Comunicación Científica) as the target-sample and we drafted a small text, copied and pasted more from the research group's website (http://ec3.ugr.es) and included several graphs and figures. Then we translated it into English using Google Translate.

At a first stage and in order to test the chances the experiment had of succeeding: this paper, written by NRG (available at http://hdl.handle.net/10481/24753) was uploaded in the end of January 2012 referencing 36 publications authored by DTS. On February 18, 2012 DTS received an email from GS Citations informing that all his papers had been cited by NRG (in Figure 1 we show the message he received for one of his papers).

FIGURE 1.E-mail alert received by the authors informing of a new (false) citation to one of their articles

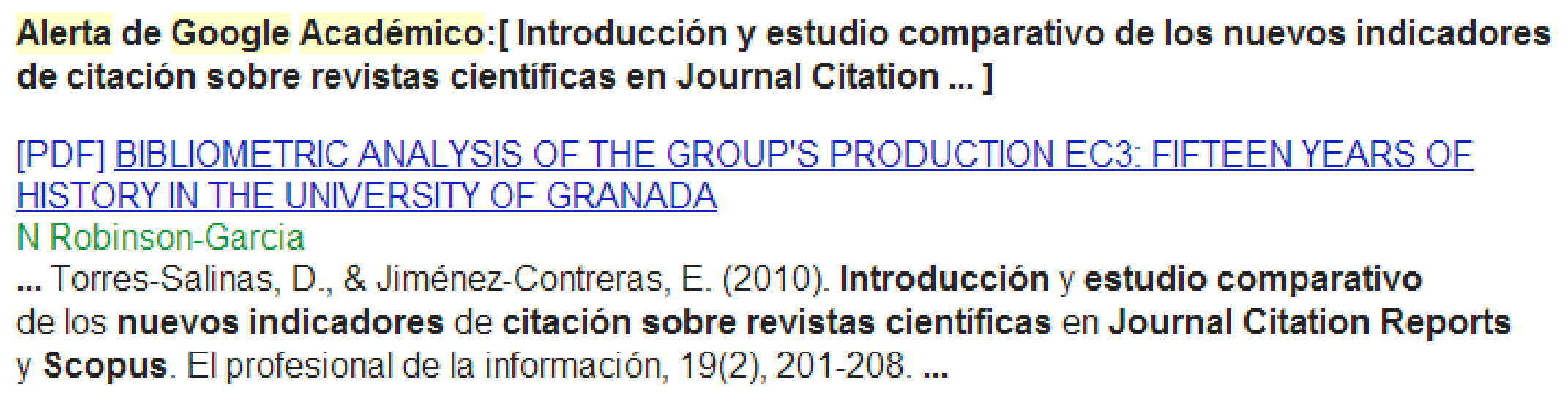

Taking into consideration the effects of this first attempt, we divided the same text into six documents. At the end of each of these documents we included references to the whole research production of the EC3 research group. In each document we preserved a similar structure to that of publications; including a title and a small abstract as well as the author. In Figure 2 we show a screenshot of the first page of some of the false papers.

FIGURE 2. Screenshot of the publications authored by M.A. Pantani-Contador

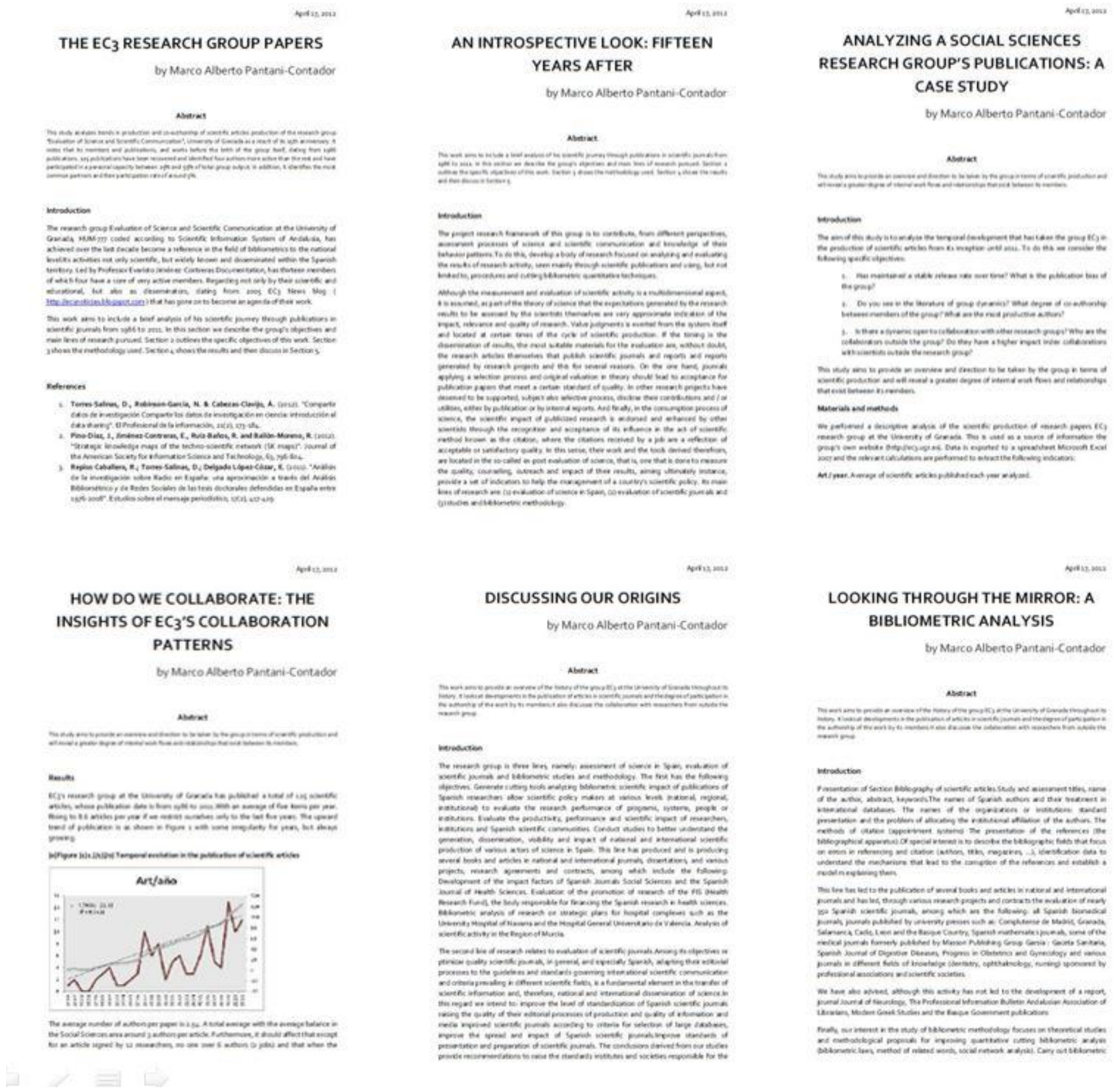

We created a false researcher named Marco Alberto Pantani-Contador, making reference to the Italian cyclist who faced many doping allegations throughout his career and the Spanish cyclist who has also been accused several times for doping. Thus, Pantani-Contador authored six documents which did not intend to be considered as published papers but simply documents made publicly available. Each document referenced 129 papers authored by at least one member of our research group. That is, we expected a total increase of 774 citations.

On 17 April, 2012 we created a webpage in html under the institutional web domain of the University of Granada (ugr.es) including references to the false papers and linking to their full-text, expecting Google Scholar to index their content. We excluded other services such as institutional or subject-based repositories as they are not obliged to undertake any bibliographic control rather than a formal one (Delgado López-Cózar & Robinson-Garcia, 2012) and we did not aim at bypassing their filters.

## Effects and consequences on the manipulation of citation data

Google indexed these documents nearly a month after they were uploaded, on May 12, 2012. At that time the members of the research group used as case study, including the authors of this paper, received an alert from GS Citations pointing out that someone called MA Pantani-Contador had cited their output (http://hdl.handle.net/10481/24753). The citation explosion was thrilling, especially in the case of the youngest researchers where their citation rates were multiplied by six, notoriously increasing in size their profiles. Table 1 shows the increase of citations the authors experienced. Obviously, the number of citations by author varies depending on the number of publications each member of the research group had as well as the inclusion of real citations received during the study period.

TABLE 1.Citations, H-Index and i10-Index values according to Google Scholar Citations before and after false citations were included

| | | Bibliometric Indicators | | |
|---|---|---|---|---|
| | | **Nr Citations** | **H-Index** | **I10-Index** |
| *Author Research Profile* | *Time Period* | *Before and After manipulation* | *Before and After manipulation* | *Before and After manipulation* |
| **Emilio Delgado López-Cózar** | **All years** | 862 → 1297 | 15 → 17 | 20 → 40 |
| | **Since 2007** | 560 → 995 | 10 → 15 | 11 → 33 |
| **Nicolás Robinson-Garcia** | **All years** | 4 → 29 | 1 → 4 | 0 → 0 |
| | **Since 2007** | 4 → 29 | 1 → 4 | 0 → 0 |
| **Daniel Torres-Salinas** | **All years** | 227 → 416 | 9 → 11 | 7 → 17 |
| | **Since 2007** | 226 → 415 | 9 → 11 | 7 → 17 |

Thus, the greatest increase is for the less-cited author, NRG, who multiplies by 7.25 the number of citations received, while DTS doubles it and EDLC experiences an increase of 1.5. We also include the variations on the H-index of each researcher. While the most significant increase is perceived in the less prolific profile, the variation for the other two is much more moderate, illustrating the irrelevance citations have to papers once they belong to the top h (Costas & Bordons, 2007). Note how in DTS' case, where the number of citations is nearly doubled, the H-index only increases by two. On the other hand, we observe how the i10-index is much more sensitive to changes. In DTS' case, the increase goes from 7 to 17, and in EDLC's case it triples for the last five years, going from 11 to 33. In Figure 1 we include a screenshot of the GS Citations profile of the one of the authors before and after Pantani-Contador's citations were included.

FIGURE 3. Screenshot of the Google Scholar Citations profile of one of the authors before and after the Google Scholar experiment took place

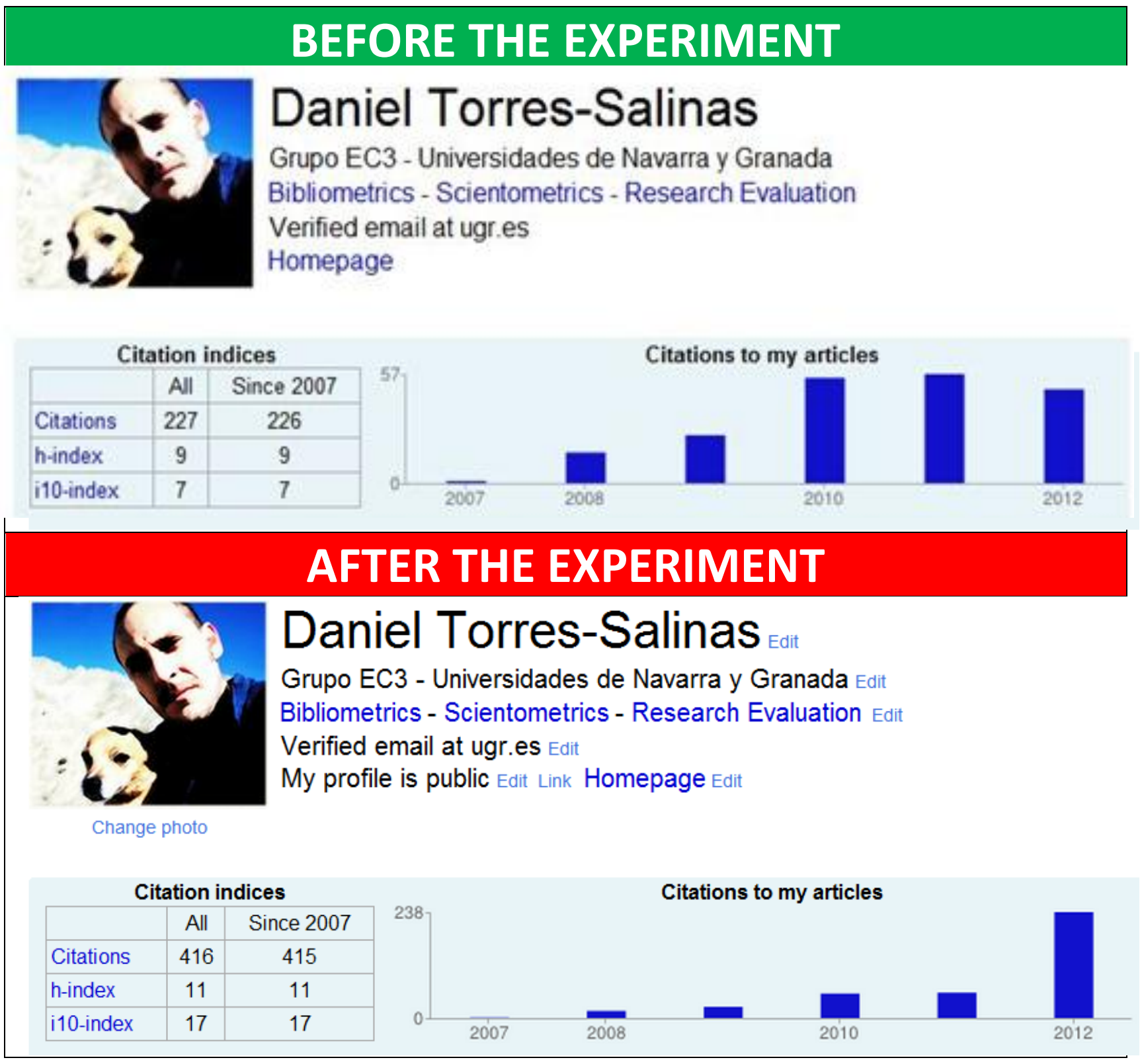

Also, it is interesting to analyze the effect this citation increase may have on the h-index for journals. For this, we have considered the two journals in which the members of the research group have published more papers and therefore, more sensitive to be manipulated. These are El Profesional de la Información with 30 papers published in this journal and Revista Española de Documentación Científica, with 33 papers. In table 2 we show the h-indexes for El Profesional de la Información and Revista Española de Documentación Científica according to Google and the increase it would have had if the citations emitted by Pantani-Contador had been included. We observe that El Profesional de la Información would be the one which would have been more influenced, as seven papers would surpass the 12 citations threshold increasing its H-index and ascending in the ranking for journals in Spanish language from position 20 to position 5. Revista Española de Documentación Científica would slightly modify its position, as only one article surpasses the 9 citations threshold that influence its h-index. Even so and due to the high number of journals with its same h-index, it would go up from position 74 to 54.

TABLE 2. Effect of the citation manipulation on journals

| Journal | h5-Index (GS Metrics) | Art > h5-Index citation threshold | Manipulated h5-Index |
|---|---|---|---|
| El Profesional de la Información | 12 | 7 | 19 |
| Revista Española de Documentación Científica | 9 | 1 | 10 |

But these are not only the journals affected. As observed in Table 3, 51 journals had their citations. In this list we find journals such as Scientometrics, with 10 papers which had an increase of 60 citations, JASIST where 3 papers increased their citations in 18 or the British Medical Journal where one paper received 6 false citations. Regardin the authors, not only the targeted group was affected by this gaming, but also all their co-authors, affecting to a total of 47 researchers.

TABLE 3. Publications affected by citation manipulation and number of increased citations by authors and journals

| Authors | Pubs | Citations | Journals | Pubs | Citations |
|---|---|---|---|---|---|
| Delgado López-Cózar E | 72 | 435 | Prof Inform | 25 | 150 |
| Jiménez Contreras, E | 58 | 348 | Rev Esp Doc Cient | 18 | 108 |
| Ruiz Pérez, R | 36 | 216 | Boletín de la Asociación Andaluza de Bibliotecarios | 11 | 66 |
| Torres Salinas, D | 32 | 189 | Scientometrics | 10 | 60 |
| Bailón Moreno, R | 18 | 108 | Rev Esp Salud Public | 4 | 24 |
| Ruiz Baños, R | 18 | 108 | Anales de Medicina Interna | 3 | 18 |
| Cabezas Clavijo, A | 8 | 48 | J Am Soc Inf Sci Tec | 3 | 18 |
| Moneda Corrochano, M | 8 | 48 | Anuario Thinkepi | 2 | 12 |
| Jurado Alameda, E | 7 | 42 | Apuntes de Ciencia y Tecnología | 2 | 12 |
| Moya Anegón, F | 7 | 42 | Aula Abierta | 2 | 12 |
| Courtial, JP | 6 | 36 | Cataloging & Classification Quarterly | 2 | 12 |
| Robinson-García, N | 4 | 25 | Ingeniería Química | 2 | 12 |
| Cordón García, JA | 4 | 24 | Knowl Organ | 2 | 12 |
| Herrera, F | 4 | 24 | Psychology Science Qu | 2 | 12 |
| Giménez Toledo, E | 3 | 18 | Revue de Bibliologie. Schémas et Schematisation | 2 | 12 |
| Moreno-Torres, JG | 3 | 18 | Anales de Documentación | 1 | 6 |
| Ferreiro Aláez, L | 2 | 12 | Arbor | 1 | 6 |
| Herrero Solana, V | 2 | 12 | Archivos de la Sociedad Española de Oftalmología | 1 | 6 |
| López-Herrera, AG | 2 | 12 | Biblio 3W, Revista Bibliográfica de Geografía y | 1 | 6 |
| | | | Ciencias Sociales | | |
| López-Huertas, MJ | 2 | 12 | Bibliodoc | 1 | 6 |
| Muñoz-Muñoz, AM | 2 | 12 | BID: Biblioteconomia y Documentació | 1 | 6 |
| Pino-Díaz, J | 2 | 12 | Boletín de la ANABAD | 1 | 6 |
| Rodríguez García, Gloria | 2 | 12 | Boletin Medes: Medicina en Español | 1 | 6 |
| Repiso Caballero, R | 2 | 12 | Brit Med J | 1 | 6 |
| Barriga, Omar A | 1 | 6 | CILIP Update | 1 | 6 |
| Bordons, M | 1 | 6 | Comunicar | 1 | 6 |
| Cobo, MJ | 1 | 6 | Educación Médica | 1 | 6 |
| Diego Carmona ML | 1 | 6 | Education for Information | 1 | 6 |
| Faba, C | 1 | 6 | Index de Enfermería | 1 | 6 |
| Fernández Cano, A | 1 | 6 | Information Research | 1 | 6 |
| Fernández, V | 1 | 6 | Int J Clin Hlth Psyc | 1 | 6 |
| Gacto Colorado, MJ | 1 | 6 | J Am Med Inform Assn | 1 | 6 |
| Guallar, J | 1 | 6 | J Doc | 1 | 6 |
| Herrera-Viedma, E | 1 | 6 | J Inf Sci | 1 | 6 |
| Liberatore, G | 1 | 6 | J Informetr | 1 | 6 |
| Marcos-Cartagena, D | 1 | 6 | J Nurs Scholarship | 1 | 6 |
| Mendoza-Parra, S | 1 | 6 | Libr Inform Sci Res | 1 | 6 |
| Moed, HF | 1 | 6 | Libri | 1 | 6 |
| Ortega, JM | 1 | 6 | Med Clin(Barc) | 1 | 6 |
| Osma Delatas, E | 1 | 6 | Nature | 1 | 6 |
| Paravic-Klijn, T | 1 | 6 | Nurs Res | 1 | 6 |
| Peis Redondo, E | 1 | 6 | Progresos de Obstetricia y Gine | 1 | 6 |
| Pérez Andrés, C | 1 | 6 | Psicothema | 1 | 6 |
| Poyatos Huertas, E | 1 | 6 | Psychology Science Quarterly | 1 | 6 |
| Roldán, A | 1 | 6 | Res Policy | 1 | 6 |
| Sanz Casado, E | 1 | 6 | Rev Esp Sanid Penit | 1 | 6 |
| Torres Ramírez, I | 1 | 6 | Rev Inves Iberoamericana Ciencia Inf Doc | 1 | 6 |
| | | | Rev Neurol | 1 | 6 |
| | | | Revista del Consejo General de Colegios de Odontólogos y Estomatólogos de España | 1 | 6 |
| | | | Revista Española de Enfermedades Digestivas | 1 | 6 |
| | | | The Lancet | 1 | 6 |

## Detection and suppression of false documents

The results exposed above were made public on May 29, 2012 in a working paper uploaded to the institutional repository of the University of Granada (Delgado López-Cózar, Robinson-García & Torres-Salinas, 2012). Simultaneously, we announced it in the research group's blog (http://ec3noticias.blogspot.com.es/2012/05/manipular-google-scholar-citations-y.html). Two days after this happened, Google erased all traces from our false researcher Pantani-Contador as well as the GS Citations profiles of the authors of this paper, which were kept to quarantine for some weeks without notifying the authors at any time and then cleaned and made publicly available. However, the initial testing document which was not reported in such working paper, still remained as caché although the pdf was no longer available in the web (see Figure 4), signifying that the suppression of Pantani-Contador and his papers was more of a reaction to our complaint than because GS had uncovered the deception. Weeks after the restitution of the authors' GS Citations profiles the false record was finally removed.

FIGURE 4. Screenshot of the citations received by an article authored by Daniel Torres-Salinas and Emilio Delgado López-Cózar

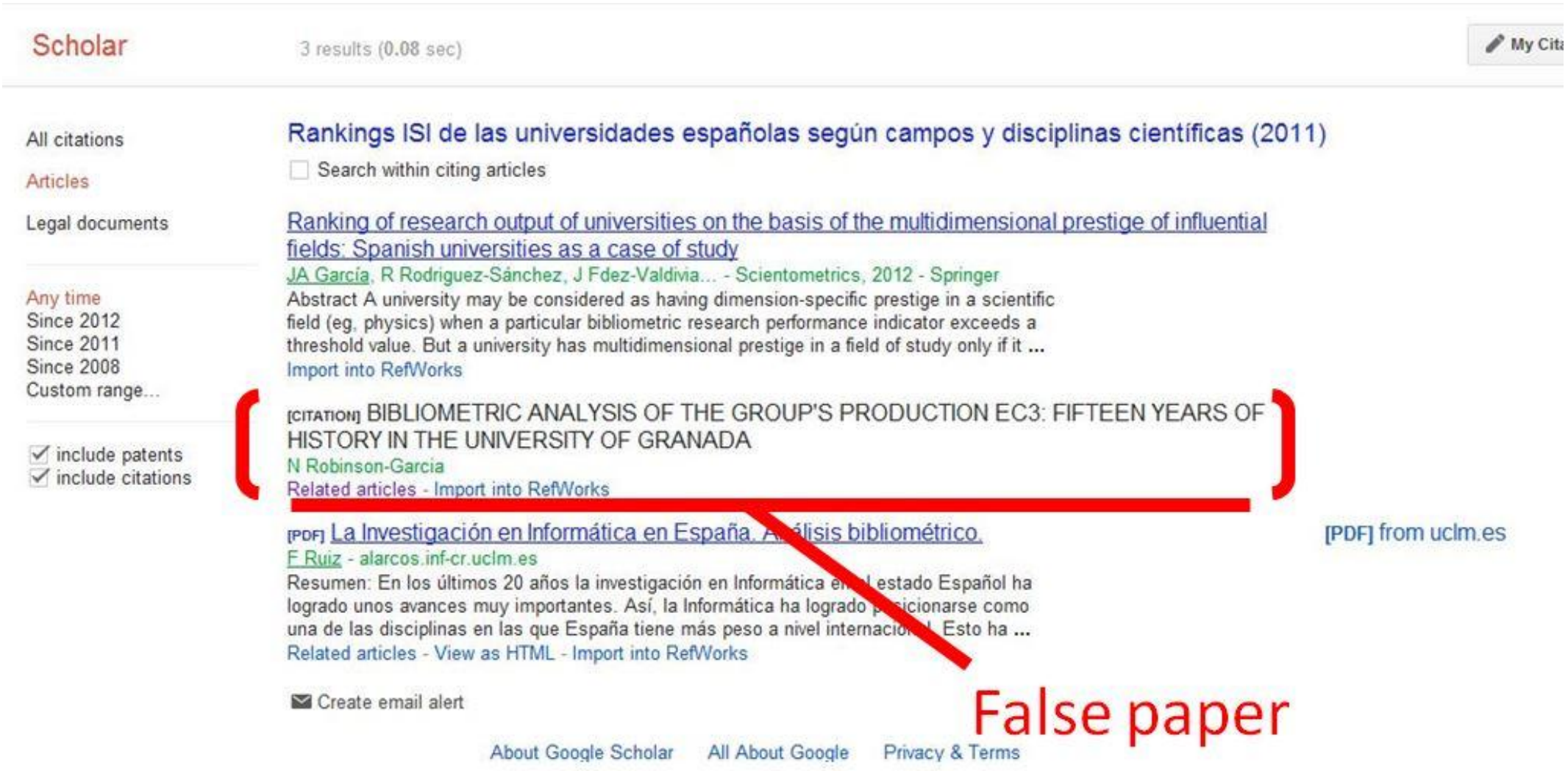

## Discussion

The results of our experiment show how easy and simple it is to modify the citation profiles offered by GS Citations and which would indirectly affect also GS Metrics. The easiness with which citation data can be manipulated raises serious concerns on the use of GS and its by-products as research evaluation tools. Also, it shows that despite its susceptibility to citation manipulation had been previously pointed out (Labbé, 2010; Beel & Gipp, 2010), nothing has been done since, transferring this limitation to GS Citations and GS Metrics. Although GS is only meant to index and retrieve all kinds of academic material in its widest sense (Meho & Yang, 2007), the inclusion of GS Citations and GS Metrics, which are evaluating tools, must include some filters and monitoring tools as well as the establishment of a set of more rigid criteria for indexing documents. Switching from a controlled environment where the output, dissemination and evaluation of scientific knowledge is monitored to an environment which lacks of any kind of control rather than researchers' consciousness is a radical novelty that encounters many dangers (Table 4).

TABLE 4. Control measures in the Web of Science vs. GS Metrics and Citations

| Web of Science | Google Scholar |
|---|---|
| - Databases select journals to be indexed | - It indexes any document belonging to an academic domain |
| - Journals select papers to be published | - Any indexed document type emits and receives citations |
| - There is a control between citing and cited documents | - It is not possible to alert on fraudulent behaviours or citation errors |
| - Fraudulent behaviours are persecuted | |

The growth of GS Metrics compared with other traditional journal rankings as well as the similarities in their results (Delgado López-Cózar & Cabezas-Clavijo, 2013) seems to validate it as good alternative. But still, it includes many methodological and technical errors (i.e., on the identification of citations, lack of transparency on the coverage or lack of quality control) converting it into an easy target to citation manipulation. This exposes the vulnerability of the product if editors and researchers are tempted to do "citations engineering" and modify their h-index by excessively self-citing their papers. Indeed, as Bartneck & Kokkelmans (2011) proved, the h-index can be easily inflated by means of three possible strategies: random self-citation, recent self-citation or, in a most refined way, sending citations only to the hot zone of their publications, that is, to those which can influence this indicator. Devious editors can easily modify their journals' h-index, having a significant effect especially for those journals with small figures on which the slightest variation can have a great impact on their ranking positioning.

When analyzing the effect of the manipulation on the indicators used by GS, that is the h-index and the i10-index, we note that many of the weaknesses appointed by others can be clearly observed (Costas & Bordons, 2007; Waltman & van Eck, 2011). In fact, the h-index seems quite stable when affecting experienced researchers while varying significantly when affecting young researchers (see table 1). Hence, the variation of the h-index is influenced by researchers' previous performance as pointed out by Waltman & van Eck (2012). As for the i10-index, it has the contrary effect, varying more significantly for experienced researchers.

## Concluding remarks

In this paper we aim at alerting on what we believe is the main shortcoming of GS Citations and Metrics: its easiness to manipulate citation counting. For this we performed a rudimentary, basic and even coarse experiment, trying to bring as much attention to it as possible. We uploaded six false documents to an institutional domain authored by a false researcher citing the scientific output of a research group. We also tried to draw as much attention as we could to it, uploading first a previous version of this paper to a repository (Delgado López-Cózar, Robinson-Garcia & Torres-Salinas, 212) in order to nourish debate among the research community and social media. This way we ensured alerting on how easy it can be to manipulate citation counting in GS and that anyone can do it, no matter how clumsy they are with technology. This means that if a more refined experiment had been done (i.e., sending citations only to those papers which would modify the h-index) it would also have gone unnoticed. Also, while using a false researcher we highlight that excluding self-citations may not be sufficient to detect such malpractices.

As suggested by many on the discussion unleashed after a previous version of this paper was made public (see e.g., Davis, 2012), one cannot expect GS to develop tools to avoid fraud when this has not been accomplished either by the traditional citation indexes (Falagas & Alexiou, 2008; Opatrný, 2008). But nevertheless, their products should be as transparent as possible so that malpractices can be easily detected. In this sense, the lack of response as well as the way Google proceeded is worrisome; as it deleted the false documents of which it had been alerted without reporting to the affected authors, remaining still the initial testing document. This lack of transparency is the main obstacle when considering GS and its by-products for research evaluation purposes. It is essential not just to display the total number of citations and h-index of researchers, but to show which of them are self-citations and the value of the h-index once these have been removed. This way we would be able to detect to which length are self-citations affecting the impact of researchers and journals. Also, GS and its by-products should include filtering systems according to document types which would avoid effects as the ones denounced in this paper. Some of these measures would be easy to adopt such as distinguishing according to document types (i.e., between journal articles, books or patents) or sources from which these are retrieved (i.e., journals, repositories or self-archiving).

However, one must acknowledge that the most efficient control or filters to avoid fraud or data manipulation are researchers' own ethical values. GS is contributing with GS Citations and GS Metrics to a democratization and popularization of research evaluation and hence, cannot avoid the responsibility of driving away the temptation to trick the metrics by means of transparency.

## Supplementary Material

All material used for the development of this study along with further evidences are available at http://hdl.handle.net/10481/24753.

## Acknowledgments

Nicolás Robinson-García is currently supported by a FPU Grant from the Ministerio de Economía y Competitividad of the Spanish Government.

## References

Aguillo, I. (2012). Is Google Scholar useful for bibliometrics? A webometric analysis. Scientometrics, 91(2), 343-351.

Bar-Ilan, J. (2008). Which h-index? - A comparison of WoS, Scopus and Google Scholar. Scientometrics, 72(2), 257-271.

Bartneck, C. & Kokkelmans, S. (2011). Detecting h-index manipulation through self-citation analysis. Scientometrics, 87(1), 85-98.

Beel, J. & Gipp, B. (2010). Academic Search Engine Spam and Google Scholar's resilience against it. Journal of Electronic Publishing, 13(3), 1-24.

Beel, J., Gipp, B. & Wilde, E. (2010). Academic Search Engine Optimization (ASEO): Optimizing scholarly literature for Google Scholar and Co. Journal of Scholarly Publishing, 41(2), 176-190.

Brown, S. & Swan, A. (2007). Researchers' use of academic libraries and their services: A report commissioned by the Research Information Network and the Consortium of Research Libraries. Available at: http://eprints.soton.ac.uk/263868/1/libraries-report-2007.pdf.

Butler, D. (2011). Computing giants launch free science metrics. Nature, 476(7358), 18.

Cabezas-Clavijo, A. & Delgado López-Cózar, E. (2013). Google Scholar e índice h en biomedicina: la popularización de la evaluación bibliométrica. Medicina Intensiva, 37(5), 343-354.

Costas, R. & Bordons, M. (2007). The h-index: Advantages, limitations and its relation with other bibliometric indicators at the micro level. Journal of Informetrics, 1(3), 193-203.

Davis, P. (2012). Gaming Google Scholar Citations, made simple and easy. Scholarly Kitchen. Available at: http://scholarlykitchen.sspnet.org/2012/12/12/gaming-google-scholar-citations-made-simple-and-easy/

Delgado López-Cózar, E. & Cabezas-Clavijo, A. (2012). Google Scholar Metrics: an unreliable tool for assessing scientific journals. El profesional de la información, 21(4), 419-427.

Delgado López-Cózar, E. & Cabezas-Clavijo, A. (2013). Ranking journals: could Google Scholar Metrics be an alternative to Journal Citation Reports and Scimago Journal Rank? Learned Publishing, 26(2), 101-113.

Delgado López-Cózar, E. & Robinson-Garcia, N. (2012). Repositories in Google Scholar Metrics or what is this document type doing in a place as such? Cybermetrics, 16(1), paper 4.

Delgado López-Cózar, E., Robinson-García, N. & Torres-Salinas, D. (2012). Manipulating Google Scholar Citations and Google Scholar Metrics: simple, easy and tempting. EC3 Working Papers 6. Available at: http://hdl.handle.net/10481/20469

Falagas, M. E. & Alexiou, V. G. (2008). The top-ten in journal impact factor manipulation. Archivum Immunologiae et Therapiae Experimentalis, 56(4), 223-226

Harzing, A.-W.& van der Wal, R. (2008). Google Scholar as a new source for citation analysis. Ethics in Science and Environmental Politics, 8, 61-73.

Harzing, A.-W. (2012). How to become an author of ESI Highly Cited Papers? Available at: http://www.harzing.com/esi_highcite.htm.

Hicks, D. (2004).The four literatures of social science. In H.F. Moed, W. Glänzel & U. Schmoch (Eds.) Handbook of quantitative science and technology research: The use of publication and patent statistics in studies of S&T systems (pp. 473-496). Netherlands: Kluwer Academic Publishers.

Jacsó, P. (2011). Google Scholar duped and deduped – the aura of ”robometrics”. Online Information Review, 35(1), 154-160.

Joint Information Systems Committee (2012). Researchers of tomorrow: The research behaviour of Generation Y doctoral students. Available at: http://www.jisc.ac.uk/media/documents/publications/reports/2012/Researchers-of-Tomorrow.pdf.

Kousha, K. & Thelwall, M. (2007). How is science cited on the Web? A classification of Google unique Web citations. Journal of the American Society for Information Science and Technology, 58(11), 1631-1644.

Kousha, K. & Thelwall, M. (2011).Assessing the citation impact of books: The role of Google Books, Google Scholar and Scopus. Journal of the American Society for Information Science and Technology, 62(11), 2147-2164.

Labbé, C. (2010). Ike Antkare one of the greatest stars in the scientific firmament. ISSI Newsletter, 6(1), 48-52.

Markland, M. (2006). Institutional repositories in the UK: What can the Google user find in there? Journal of Librarianship and Information Science, 38(4), 221-228.

Meho, L. & Yang, K. (2007). Impact of data sources on citation counts and rankings of LIS faculty: Web of Science versus Scopus and Google Scholar. Journal of the American Society for Information Science and Technology, 58(13), 2105-2125.

Nicholas, D., Clark, D., Rowlands, I. & Jamali, H.R. (2009). Online use and information seeking behaviour: Institutional and subject comparisons of UK researchers. Journal of Information Science, 35(6), 660-676.

Opatrný, T. (2008). Playing the system to give low-impact journal more clout.Nature, 455, 167.

Oransky, I. (2012). A first? Papers retracted for citation manipulation. Retraction Watch. Available at: http://retractionwatch.wordpress.com/2012/07/05/a-first-papers-retracted-for-citation-manipulation/

Ortega, J.L. & Aguillo, I. (2013). Institutional and country collaboration in an online service of scientific profiles: Google Scholar Citations. Journal of Informetrics, 7(2), 394-403.

Peters, D.P. & Ceci, S.J. (1982). Peer-review practices of psychological journals – the fate of accepted, published articles, submitted again. Behavioral and Brain Sciences, 5(2), 187-255.

Sokal, A. (1996). A physicist experiments with cultural studies. Lingua Franca, 6(4), 62-64.

Torres-Salinas, D., Ruiz-Pérez, R. & Delgado López-Cózar, E. (2009). Google Scholar como herramienta para la evaluación científica. El profesional de la información, 18(5), 501-510.

Walters, W. (2011).Comparative recall and precision of simple and expert searches in Google Scholar and eight other databases. portal: Libraries and the Academy, 11(4), 971-1006.

Waltman, L. & van Eck, N.J. (2012). The inconsistency of the h-index. Journal of the American Society for Information Science and Technology, 63(2), 406-415.

Wouters, P. & Costas, R. (2012). Users, narcissism and control – tracking the impact of scholarly publications in the 21st century. Available at: http://www.surf.nl/nl/publicaties/Documents/Users%20narcissism%20and%20control.pdf.